\documentclass[12pt]{article}
\usepackage{amssymb,amsmath,graphicx}
\usepackage{longtable}
\usepackage{hyperref}
                               
\newcommand{\tab}{\hspace{5mm}}

\begin{document}

\begin{center}
\textbf{{\Large ON AN IMPROVEMENT OF THE
\vspace{0.2 cm}
PLANCK RADIATION ENERGY\\
\vspace{0.2 cm}
DISTRIBUTION}}\\
\vspace{0.5 cm}
{Diego Sa\'{a}}
\footnote{Escuela Polit\'{e}cnica Nacional. Quito --Ecuador. email: dsaa@server.epn.edu.ec}\\
Copyright {\copyright}2006\\
\end{center}

\begin{abstract}
{The probability distribution function for thermodynamics and econophysics 
is obtained by solving an equilibrium equation. This approach 
is different from the common one of optimizing the entropy of 
the system or obtaining the state of maximum probability, which 
usually obtains as a result the Boltzmann distribution.  
The Gamma distribution is proposed as a better equation to describe the blackbody radiation in substitution of Planck's radiation equation. Also, a new form of entropy is proposed, that maintains the correct relation with the Clausius' formula.}
\end{abstract}


\textit{keywords}: statistical mechanics, thermodynamics, econophysics, thermodynamic equilibrium, probability distributions\\

\textbf{1. INTRODUCTION }\\

Kirchhoff introduced the concept of a perfectly black body, being 
one which absorbs all incident radiation. He was the first in 
presenting the enunciation of the principle of the balance between 
emission and absorption around 1860. The principle of equilibrium 
of emission and absorption between different agents is, essentially, 
the method that will be used in the present paper to develop 
a new equation that describes the radiation distribution of a 
black body.

By 1900 Planck devised the known equation that describes the 
distribution of radiation emitted by a black body, based on hunch 
and pure guessing. Although it is rather accurate over the entire 
range of frequencies it is suggested here that it is not appropriate to describe the blackbody radiation distribution.

``Although the introduction of energy quanta a century 
ago has led to the currently accepted explanation within quantum 
theory, there is still no firm conclusion as to whether or not 
blackbody radiation can be explained within classical physics.'' 
\cite{boyer}

One of the basic assumptions in this paper is that energy exists 
in packets, or quanta, which have continuous values.

This deserves some explanation because this is against one of 
the most basic and accepted tenets of Physics, in the so-called 
area of quantum theory. 

The elementary particles called fermions, such as the electron 
and proton, have some very specific amounts of energy and also 
the orbits of the electrons in the atoms seem to be at discrete 
levels. Consequently, it can be safely accepted that those elementary 
particles satisfy the classical quantum theory. Moreover, it 
can be accepted that, when the electrons jump between levels 
in the atomic orbits, they emit and absorb photons with discrete levels of energy.

However, this discrete character is not logically and necessarily extrapolated to the photons produced, for example, in the blackbody or an antenna radiation. It has not been proved that such photons have discrete levels of energy. On the contrary, electromagnetic radiation seems to have a continuous spectrum in the whole range of frequencies or wavelengths. The present author has not been able to find any experiment supporting the opposite position. \\

It seems that some physicists might have some misconceptions about the current blackbody radiation theory. For example, the peer reviewers of Physica A, under the orders of Prof. C. Tsallis, used the argument that ``It is obvious that the energy spectrum given by the Planck distribution is continuous'', to reject a former version of the present paper.

On the light of the points of view suggested by the following reference, such argument gives the impression of being wrong. 

Let us refer as ``A\&F'' to the book of Alonso \& Finn, ``\textit{Fundamental University Physics, Volume III, Quantum and Statistical Mechanics}'', Addison-Wesley, 1968, which is referred to in the present paper as \cite{alonso}.
 
By definition, Quantum Mechanics deals with \textit{energy levels}. This is suggested in Fig. 2-10 of A\&F. 

In section 10-3, A\&F carry out the proof of the ``Maxwell-Boltzmann distribution.'' In particular, in the second paragraph of that section, the authors mention ``\textit{the number of particles present in each of the energy levels}''. Fig.10-1, appearing in the same section, is also a good illustration of this concept. The derivation of the ``probability distribution according to the Maxwell-Boltzmann distribution'', equation (10.8), is very clear and should be reviewed to confirm that the authors A\&F always use \textit{discrete energy levels}. After using a simple mathematical optimization technique they find the ``partition of maximum probability'' and the law of ``Maxwell-Boltzmann distribution'', equations (10.9) and (10.11), which provide the integer number of particles in each energy level.

It is obvious that the products of integer numbers by discrete energy levels produce discrete levels for the energy distribution.

More relevant to the present paper is the derivation of the \textit{Planck radiation energy distribution}, eq. (13.20), which is carried out by A\&F in section 13.6, that they call ``maybe the most important application of the Bose-Einstein statistics'' (rather similar to the Maxwell-Boltzmann distribution, given the additional considerations identified in the cited book). The authors mention that it is only an approximation that the energy spectrum could be considered continuous, under the condition that ``the cavity be great relative to the mean radiation wavelength \textit{because the energy difference among two successive energy levels is small}.'' 
This last condition is satisfied because, in general, we have a ``big box'', as pictured in Fig. 2-10 (b), compared with the small values obtained for energy, which is just the product of the frequencies of the packets by the Planck constant. But this does not mean that Quantum Mechanics does not consider energy levels or that those levels have somehow disappeared during the derivation.\\

Moreover, the whole mentioned derivation produces the mathematical forms of the Planck radiation distributions, which are different from the distribution obtained under the assumption of continuous energy values for the photons, as is proved in this paper.\\

Some investigators have recognized the fact that there are two different maxima for the Planck function, which they justify using explanations that the present author and even their students consider implausible. Read, for example, the anecdotes after equations (1.24) and (1.25) of section 1.2.1 in the book by Bohren \& Clothiaux \cite{bohren}. Those authors conclude that ``Failure to recognize that the maximum of a distribution function depends on how it is plotted has led and no doubt will continue to lead to errors.'' On the other hand, Soffer \& Lynch \cite{soffer} direct their quest to the spectral sensitivity of the human eye in order to unify (within the eye) the different maxima, with the premise that the problem ``results from the choice of units in which the solar spectrum is plotted.''

The same or a very similar argument seems to be getting out of control with other investigators \cite{heald},\cite{overduin} that write a variety of equations that supposedly make an equivalent ``bookkeeping'' of the ``Wien peak''. For example \cite{heald} reports at least five equations and refers to four of them saying that ``[a]ll four of these dispersion rules, with their corresponding graphs and formulas, represent the same physical spectrum.'' The same author, making reference to a publication of the American Association of Physics Teachers that reports quibbles, misunderstandings and mistakes in Physics, says that ``The analysis and interpretation of continuous spectra has a reputation for leading people astray''. However, the physicists and the Physics teachers in particular would not move a finger to correct those mistakes. The attempt to publish a former version of the present paper received the following reply, supposedly stated in the
Statement of Editorial Policy: ``AJP [a publication of the American Association of Physics Teachers] may only 
consider for publication manuscripts dealing with concepts accepted by the general community of physicists''.

There is no doubt that the current concepts about blackbody radiation, Planck equations, Wien displacement law, Boltzmann entropy and others that are challenged in the present paper are unanimously accepted by the scientific community. Consequently, the absolute acceptance of those (and other) concepts puts an unsurmountable barrier to the possibility of publishing a paper that looks disident, even if it could be correct.\\

I consider that fermions, such as the electrons, satisfy the Quantum postulates because they are packets with very specific energy levels, which amounts to say that they have some very specific frequencies and wavelengths. For example, Millikan's experiment detected only discrete levels of energy, corresponding to the finite number of electrons in the oil drops.

On the other side, I consider that the photons (and in general bosons) really do not have discrete energy levels. They are energy quanta (in the sense that they are packets) that can have any positive continuous frequency. \\

It is a suspicion of the present author that Planck was misled by some incorrect experimental results by H. Rubens, who visited Planck in October 1900, and others. As Planck himself notes: ``Whilst for small values of the energy and for short waves, Wien's law was satisfactorily confirmed, noteworthy deviations for larger wavelengths were found, first by O. Lummer and E. Pringsheim, and finally by H. Rubens and F. Kurlbaum'' \cite{planck}. See, for example, the strange curves transcribed as figure 2 in the paper by Abraham Pais \cite{pais} and attributed to Rubens\&Kurlbaum. All the curves (Wien, Rayleigh, Thiesen, Lummer and the observed values) are clamped around two fixed points at 0 \ensuremath{^\circ}C and 1000 \ensuremath{^\circ}C, with a difference of up to 30\% between the observed values and the Wien curve at around 500 \ensuremath{^\circ}C. Besides, nobody would accept by now that the observed values are much better modeled by the Rayleigh-Jeans than by the Wien's equation, as shown in such figure.

The ``satisfactory confirmation'' of Wien's law was given, evidently, by the careful measurements, ``very good ones'' \cite{pais}, by Friedrich Paschen, who presented contributions to the Prusian Academy of Sciences between 1987 and 1899 that showed a ``perfect agreement with Wien's theory'' \cite{barrachina}.

As Planck felt the need to take into account those new experimental results and ``improve'' Wien's law, ``Planck interpolated between the laws of Wien and Rayleigh-Jeans but found that he could only derive a satisfactory equation using what he thought was merely a mathematical trick, namely that light is only emitted in \textit{packets}. He did not 
for many years believe that these packets, known as \textit{quanta}, corresponded with reality.'' \cite{wiki}\\

Nothwithstanding the fact that quantum theory, a theory generally accepted by now, was originated in Planck's theory, it will be suggested in this paper that the Planck's blackbody radiation equations might be incorrect.\\

\textbf{2. THE PLANCK'S EQUATIONS}\\

The temperature of a body is a measurable effect furnished by 
the statistical average of the energies of the enormous amount 
of photons emitted by that body. However, this is not the practical 
way to compute the temperature because we don't know the temperature of the individual photons.

I have two objectives in this and the following section: first, summarize some 
of the equations used in statistical thermodynamics and, second, 
attempt to put in evidence that the temperature, of a blackbody 
in particular but in general the temperature of any body, can 
profitably be interpreted in Physics as a certain frequency of 
the photons emitted by that body.

First, the energy density radiated by a black body per unit volume, for a given wavelength $\lambda $ and absolute temperature \textit{T}, is expressed by Planck's radiation 
formula:
\begin{equation}
\rho (\lambda ,T)=\frac{8\cdot \pi \cdot h\cdot c}{\lambda ^{5} } \cdot
\frac{1}{\exp \left( \frac{h\cdot c}{\lambda \cdot k\cdot T} \right) -1}
\left[ \frac{joule}{m^{4} } \right]
\end{equation}

where \textit{k} is the Boltzmann constant $k=1.3805\cdot 10^{-23} \cdot joule/K$)

\textit{h} is Planck's constant and \textit{c} is the speed of light in 
vacuum. \\

The total energy radiated by the body at a given temperature 
is given by Stefan-Boltzmann law, which is obtained integrating 
the previous equation (1) for all values of wavelength from zero 
to infinity:

\begin{equation}
E=a\cdot T^{4}
\end{equation}

where \textit{a} is a constant given, in terms of more basic constants, 
by:

\begin{equation}
a=\frac{8\cdot \pi ^{5} \cdot k^{4} }{15\cdot h^{3} \cdot c^{3} }
\end{equation}

The numerical value of this constant is: $a=7.562473\cdot 10^{-16} \cdot \frac{joule}{m^{3} \cdot K^{4} } $ \cite{alonso}\\

The same resulting equation (2) and constant ``\textit{a}'' 
are also obtained by integrating, for all values of the frequency, 
from zero to infinity, the expression of Planck's radiation formula 
expressed in terms of frequency instead of wavelength, which 
is:

\begin{equation}
\rho (\nu ,T)=\frac{8\cdot \pi \cdot h\cdot \nu ^{3} }{c^{3} } \cdot
\frac{1}{\exp \left( \frac{h\cdot \nu }{k\cdot T} \right) -1} \left[
\frac{joule\cdot s}{m^{3} } \right]
\end{equation}

Let us call \textit{maxnu} the argument of the exponential function 
that appears in this equation:

\begin{equation}
maxnu=\frac{h\cdot \nu }{k\cdot T}
\end{equation}

Let us replace this definition in (4) and solve for \textit{maxnu} 
for the point at which equation (4) predicts the maximum radiation per unit frequency. 
The maximum is obtained when \textit{maxnu} is equal to 3*(1-exp(-\textit{maxnu})). 
That is, when \textit{maxnu} has the value 2.821439\dots \\

On the other hand, if we call \textit{maxlambda} the argument of the 
exponential function of equation (1):

\begin{equation}
maxlambda=\frac{h\cdot c}{k\cdot T\cdot \lambda } 
\end{equation}

Then, the point of maximum radiation per unit wavelength, computed from equation (1) by replacing in it this definition (6), is obtained when \textit{maxlambda} is equal to 5*(1-exp(-\textit{maxlambda})). That is, when \textit{maxlambda} has the value 4.965114\dots \\

Let us remember the well-known relation between the wavelength 
and frequency of a given photon, which says that its product 
is equal to the speed of light:
\begin{equation}
\lambda \cdot \nu =c
\end{equation}

if $c/\lambda$ from this expression is replaced by $\nu$ in the right hand side of equation (6) then both variables, \textit{maxnu} and \textit{maxlambda} 
are proved to be equal to exactly the same common symbolic expression. 
However, the said common expression does not have a unique numerical value, but two values, given respectively by the variables \textit{maxnu} and \textit{maxlambda} computed above.

If we were expecting, intuitively, that the frequency and 
wavelength of the photons emitted at the point of maximum radiation 
density should be the same, whether they are computed with equations 
(5) or (6), this proves to be wrong.

The reason why these two constants, \textit{maxnu} and \textit{maxlambda}, 
are numerically different in current Physics is because 
there are two kinds of photons that participate in the computation 
of each one of these variables. So, the problem remains as to 
which one of them, both, or none, is \textit{the} correct photon emitted 
at the physical point of maximum radiation. As was said before, 
the formula (6) is usually used to compute the properties of 
the photons, but with no clear explanation as to why the different 
value produced with equation (5) is incorrect or does not correspond 
to the same maximum.\\

In principle, either of the last two relations, (5) or (6), could 
be used to compute the values of the frequency or wavelength 
of the photons emitted when the radiation has its maximum value. 
Namely, solving (6) for wavelength:
\begin{equation}
\lambda max=\frac{1}{maxlambda\cdot T} \cdot c\cdot \frac{h}{k} 
\end{equation}

Consequently, from this equation, we obtain:

\begin{equation}
\lambda max\cdot T=b
\end{equation}

This is Wien's law of displacement that allows to compute the 
wavelength of the photons emitted at the point where the density 
of radiation per unit wavelength is a maximum, by dividing the constant \textit{b} by the absolute temperature of the body. \\

\textit{b} is called Wien's displacement constant and has units of 
length multiplied by temperature:

\begin{equation}
b=\frac{h\cdot c}{k\cdot maxlambda} =0.0028980841\cdot m\cdot K
\nonumber
\end{equation}

On the other hand, solving equation (5) for the frequency, we 
find an unnamed equation that is never used to compute the frequency 
of the same photons:

\begin{equation}
\nu max=maxnu\cdot T\cdot \frac{k}{h} 
\end{equation}

Ok, not \textit{never} but at least \textit{seldom} used, because equation 
(8), which is derived from equation (6), is the most commonly 
used equation. However, as examples of the use of equation 
(10), which is derived from equation (5), see \cite{blumler},  \cite{armitage}, \cite{fitzpatrick}, \cite{kaiser}, \cite{opps}, \cite{strong}.

The results obtained with these equations are compared in the following section.

Obviously, the frequency and wavelength provided by equations 
(8) and (10), do not belong to the same kind of photons, emitted 
at the point of maximum radiation.\\

\textbf{3. THE TEMPERATURE IS GIVEN BY THE FREQUENCY OF PHOTONS}\\

One of the strongest hints, to accept as true that the temperature 
is frequency is obtained by replacing the wavelength 
from Wien's law, equation (9), in equation (7) and solving for \textit{T}:

\begin{equation}
T=\nu max\cdot \frac{b}{c}
\end{equation}

That is, the temperature of a body is equal to a frequency multiplied 
by a constant that simply converts units. Usually we know the 
temperature and wish to know the frequency, so let us define 
the constant \textit{kel} to convert the temperature in Kelvin degrees 
to Hertz as:

\begin{equation}
kel=\frac{c}{b} =1.0344505\cdot 10^{11} \cdot \frac{Hz}{K}
\nonumber
\end{equation}

Once the frequency is computed, multiplying this constant by 
the temperature of the body, we are able to solve equation (7) 
for wavelength. Alternatively, divide the constant \textit{b} by the 
temperature to obtain wavelength.\\

For example, to a temperature 
of 5530 \ensuremath{^\circ}K (temperature of the sun, computed by Wien in his Nobel lecture \cite{wien}), correspond photons with wavelength of visible light 
of about 524 nm (Wien reports 532 nm. Verify in \cite{dunlap} that this corresponds to a sun of green color, assuming that the photons emitted at the point of maximum radiation have some relation with the sun's color) and frequency around 5.72*10$^{14}$ Hz. Compare this with the value of 3.25*10$^{14}$ Hz obtained with the use of equation (10).

The product of wavelength and frequency of the two kinds of photons for which the radiation is a maximum (obtained from the two density of radiation equations, (8) and 
(10) ) is about 0.568 of the speed of light (try for example multiplying the values reported in \cite{israel}, \cite{kaiser}), which is the same proportion as between \textit{maxnu} and  \textit{maxlambda}. This proves that they are not the same kind of photon. \\

More evident and dramatic becomes the difference if we solve equation (10) for the temperature that corresponds to the wavelength used by Wien, 532 nm. This gives around 9586 \ensuremath{^\circ}K. This means that there is a difference of more than 4000 \ensuremath{^\circ}K between the sun's temperatures predicted by two supposedly valid equations of current thermodynamics. I wonder how the physicists justify this sort of inconsistencies or are blind to them. In the particular case of the sun's temperature, for example, why is it usually computed with equation (8) instead of with equation (10)?. Several references, including the same Wien \cite{wien}, insist that the point of maximum radiation of the sun is somewhere in the visible range, mainly between 475 and 532 nm, which corresponds to a sun of blue, turquoise or green color!... physicists have a lot of things to explain. Nevertheless, if we assume as known the sun's temperature of 5530 \ensuremath{^\circ}K and compute the corresponding wavelength for the maximum, again with the help of equations (10) and (7), we find 922 nm, which is in the infrared, well outside the visible range.

It should be noted that, in practice, neither of these equations seems to be used. For example in \cite{marks} the authors report that equation $\lambda T = 3670 \mu m\cdot K$ is a ``more useful displacement law than Wien's'' (without explaining why it is more useful or how it is obtained). This equation is far away from both the $2898.08 \mu m\cdot K$, reported earlier in this paper, and the $5100 \mu m\cdot K$, produced respectively by equations (9) and (10), but much closer, within 2\%, to the value that can be computed with the theory proposed in the following:
$\lambda \cdot T = h\cdot c/4\cdot k = 3597.33 \mu m\cdot K$. With this equation and with the more updated temperature for the sun of around 6000 \ensuremath{^\circ}K we are able to compute a wavelength of 600 nm, which corresponds to a nice orange sun (see again \cite{dunlap}).\\

An interesting application of the constant \textit{kel} converts Boltzmann's constant into Planck's constant. This means 
that Boltzmann's constant (by the way, never written as such 
by Boltzmann but by Planck, and whose currently accepted value might be a little greater than needed for the theory being developed here) is redundant and superfluous if we 
use only frequencies instead of temperatures in our computations.\\

\textbf{4. DEVELOPMENT OF A NEW EQUATION FOR RADIATION}

The criteria used to develop a new equation for radiation, which 
hopefully will provide an expression for the radiation law that 
avoids the above-mentioned inconsistencies, are the following:\\

a.- The product of wavelength by frequency for a certain kind 
of photons should give the speed of light.\\

b.- The point of the maximum radiation, expressed either in terms 
of wavelength or of frequency should be the same. \\

c.- The photon energy is assumed to have a continuous spectrum. 
This means that the photon energy is quantized but does not have discrete values.\\

The conditions ``a'' and ``b'' are clear 
and intuitive and the last one was explained in the introduction
in the sense that the physical particles show quantum behavior; this means that their ``size'' (wavelength) 
and energy (frequencies) are of some fixed values. But the experimentalists 
have not been able to find discrete values for the corresponding 
photon variables. Rather, the values of photon energy and wavelength 
have a seemingly infinite and continuous range.\\

In an open system, the radiation process emits photons to the 
space, which are not absorbed again by the system, so the mass 
or temperature of the system is reduced. \\
But what we will be considering in the following is the case 
of a closed system, which is in thermal equilibrium.\\

Consider an individual photon emitted by a black body. This photon 
bumps with other particles and loses and gains energy in this 
process, before it is completely absorbed or eventually gets 
its opportunity to get out through the small window of the blackbody for its energy to be measured by us. In fact this opportunity 
is so small that we are able to consider the blackbody as a 
closed system in thermodynamic equilibrium.

\tab Let us try to determine what the distribution of the energies 
of the individual photons is. \\

The statistical weight (number of possible states of a system) 
of a given (discrete) distribution is
\begin{equation}
\Omega =\frac{N!}{\prod _{i} n_{i} !} 
\end{equation}

When the distribution is continuous, each partition \textit{i} has 
only one particle in it and hence each factorial in the denominator 
is the unity. Moreover, the factorial in the numerator must be 
interpreted as the Gamma function, which is the continuous equivalent 
of the factorial.

Consequently, the entropy \textit{S}, that is just the statistical 
weight $\Omega $, is defined as
\begin{equation}
S=\Omega =\Gamma (N)=\Gamma (p)
\end{equation}
where the change from \textit{N} to the new parameter \textit{p} is to try to understand that now we have continuous and not integer values.\\
In the following we will need to study the actual entropy distribution of these states among the different values of a certain variable, \textit{x}, such as frequency, and not merely the total number of states. Therefore, the appropriate form of the entropy will be:
\begin{equation}
S(x)=\Gamma (p,0,x)
\end{equation}
This provides the cumulative number of states for the variable between 0 and some value \textit{x}. The cumulative normalized entropy becomes:
\begin{equation}
S(x)=\frac{\Gamma (p,0,x)}{\Gamma (p)}=\frac{\Gamma (p)-\Gamma (p,x)}{\Gamma (p)}=1-\frac{\Gamma (p,x)}{\Gamma (p)}
\end{equation}
Also, to convert the units of the variable \textit{x} to a non-dimensional form, we include a constant, $\lambda$:
\begin{equation}
S(x)=1-\frac{\Gamma (p, \lambda \cdot x)}{\Gamma (p)}
\end{equation}

We assume that a system in a state of equilibrium is changing 
between microcanonical states in a reversible form. In such case, 
remember that for reversible processes the change of entropy is given, by the Clausius' formula, as proportional to the change of heat:
\begin{equation}
dS(x)=\frac{dQ(x)}{T} 
\end{equation}

where \textit{T} is the average temperature of the system.\\

Consequently, replacing (16) in (17) and taking the derivative, we obtain the equation that 
describes the probability distribution of energy in a continuous system, as is assumed the blackbody. More specifically it also describes the specific heat of compounds, where the low-temperature zone corresponding to the so-called \textit{Schottky anomaly} has this distribution:

\begin{equation}
\frac {dQ(x)}{dx}=T\cdot \frac{dS(x)}{dx}=T\cdot \frac{\lambda ^{p} \cdot x^{p-1} }{Exp(\lambda \cdot x)\cdot \Gamma (p)}
\end{equation}

The previous is the most simple derivation, which proves that the energy distribution per unit \textit{x} is a Gamma distribution.\\

An easier and profitable understanding of the following section is achieved if we interpret the energy distribution as a probability distribution function (pdf) and the entropy as the corresponding cumulative distribution function (cdf). It should be noted that ``according to Boltzmann, entropy is a measure for physical probability'' \cite{planck}. \\

\textbf{5. DERIVATION FROM FIRST PRINCIPLES}\\

Let us assume a \textit{continuous} random variable ``\textit{X}'' 
that represents some physical quantity such as the amount of 
money or energy possessed by agents or actors in a certain system. \\

The probability that \textit{X} lies in a small interval (x, x+$\Delta x$) is
\begin{equation}
P\lbrack x<X\leq x+\Delta x]=F(x+\Delta x)-F(x)=\frac{F(x+\Delta
x)-F(x)}{\Delta x} \Delta x
\end{equation}

If $\Delta x$ is small and \textit{F}(\textit{x}) is differentiable then the following approximation holds:
\begin{equation}
P\lbrack x<X\leq x+\Delta x]=F'(x)\Delta x
\end{equation}

in which the prime denotes the derivative of \textit{F}(\textit{x}). Whenever it exists, the derivative of \textit{F}(\textit{x}) is called the \textit{probability density function} (pdf) of ``\textit{X}'' and is denoted by \textit{f}(\textit{x}). Thus,

\begin{equation}
f(x)=\frac{dF(x)}{dx} 
\end{equation}

The units of the pdf are inverse to those of the random variable \textit{X}.\\
Therefore, given the pdf, the distribution function of \textit{X} 
is computable as

\begin{equation}
F(x)=\int\nolimits_{-\infty }^{x}f(u)du .
\end{equation}

This is the probability that \textit{X} is less than \textit{x}. \textit{F}(\textit{x}) 
is called the \textit{cumulative distribution function} (cdf):

\begin{equation}
P\lbrack -\infty <X\leq x]=F(x)
\end{equation}

Since the cdf is a non-decreasing function, the pdf is always 
nonnegative:

\begin{equation}
f(x)\geq 0
\end{equation}

In the following we are going to argue that, in a model in which 
the agents interact exchanging the quantity \textit{X}, the equilibrium 
distribution of this variable is a Gamma distribution. \\

We define a kinetic continuous model in which, for facility of 
analysis, consider a finite number \textit{N} of agents that interact 
with each other by ``trading'' or ``colliding'' 
between them, and exchanging random amounts of the quantity \textit{X}, 
generally referred to as ``money'' (or energy) in the following. \\

The evolution of the system is then carried out in the following 
way. At every time step two agents \textit{i} and \textit{j} are extracted randomly 
and an amount of money $\Delta x$ is exchanged between them,
\begin{eqnarray}
{x'_{i} = x_{i} - \Delta x,}
\nonumber\\
{x'_{j} = x_{j} + \Delta x.}
\end{eqnarray}

It can be noticed that in this way the quantity \textit{X} is conserved 
after each transaction, $x'_{i} + x'_{j} = x_{i} + x_{j}$, where $x'_{i}$ and $x'_{j}$ are the agent's wealth after 
the transaction has taken place.\\

The proportion of the variable \textit{x} (money) owned by the system 
at a certain point \textit{x} is: 
\begin{equation}
x f(x)
\nonumber
\end{equation}

Then, the amount of money, \textit{a}, corresponding to a small interval 
$\Delta x$ around \textit{x}, is computed as the product of \textit{x} by the probability density of occurrence of such amount at that point and by the width of the interval:

\begin{equation}
a = x f(x) \Delta x
\end{equation}

Similarly, for the point \textit{x}+$\Delta x$:

\begin{equation}
b = (x + \Delta x) f(x + \Delta x) \Delta x
\end{equation}

Let us call \textit{delx} the amount of money given by the agent at (\textit{x}+ $\Delta x$) to the agent at \textit{x}. \\

If the system is in a state of statistical equilibrium, the amount 
of money possessed by the agent at \textit{x} after the transaction, let us call \textit{a}', must be equal to the amount of money possessed by the (wealthier) agent at $(\textit{x}+ \Delta x)$ before the transaction, that is \textit{b}, minus the amount given by this agent in the transaction, that is \textit{delx}; and vice versa, \textit{b}' is equal to the amount of money possessed by the agent at \textit{x}, that is \textit{a}, plus the amount received from the other agent, that is \textit{delx}. In the form of equations this means:
\begin{eqnarray}
a' = b - delx
\\
b' = a + delx
\end{eqnarray}

But, at equilibrium, the amounts of money before and after the 
transaction must be equal, therefore \textit{a}' = \textit{a} and \textit{b}' 
= \textit{b}, so the two previous equations are in fact the same one.\\
These equations reflect the fact that in a one-dimensional, completely elastic, head-on, collision, classical mechanics predicts that all the energy of one of the agents is transferred to the other agent and vice versa. The amount of ``saving'' in the collisions is dependent, in general, on the number of degrees of freedom.\\

If we assume that there is a first agent at position \textit{x} that is in equilibrium with the second consecutive element at (\textit{x}+ $\Delta x$), and if this second element is in equilibrium with the third, then the first is also in equilibrium with the third and thus, successively, we conclude that two agents at any distance will be in equilibrium.\\

Next, the amount of money traded between these two agents is 
assumed to be composed of two terms. \\

The first term is proportional to the difference between the 
amounts of money possessed by the two agents, $\Delta F(x)$:

\begin{equation}
p \; \Delta F(x) = p \;f(x) \, \Delta x
\end{equation}

Where \textit{p} is the constant of proportionality.
This equality is justified above: the derivative of the probability 
distribution of a variable is equal to its pdf. \\

Actually \textit{F}(\textit{x}) is probability; hence, in order to fix 
the units it must be multiplied by the amount \textit{M} of total 
money of the system, because the variable interchanged in the 
process is assumed to be money; however, this amount can be seen 
that is simplified in the following equations.\\

The physical interpretation of equation (30) and the objective 
of all this exercise is the following: if both agents have the 
same amount of money then the difference between these amounts 
is zero and there is no money interchange since the system would 
be in equilibrium. But if we think that one of the agents (e.g. 
a photon) has more money (energy) then it can give more money 
in each transaction (energy in a collision), in proportion to 
its greater wealth (energy difference).\\

In the previous expression, the constant of proportionality, \textit{ p}, is a function of some given positive numbers usually extracted from a uniform distribution between 0 and 1. Since the number of transactions is very large we can safely assume that the net effect, if \textit{p} is a random variable instead of a constant, is equivalent to its average. \\

The second term is the amount ``saved'' by the wealthier 
agent and is proportional to the absolute amount of money possessed by that agent: 
\begin{equation}
\lambda  b 
\end{equation}
Or, expanding \textit{b}
\begin{equation}
\lambda (x + \Delta x) f(x + \Delta x ) \Delta x
\end{equation}

where the variable $\lambda $ is a function of the saving propensity of the agent and has inverse units to those of the variable \textit{x}. \\

In summary the amount of money given by the wealthier agent to 
the poorer one is
\begin{equation}
delx = (p \; f(x) \Delta x - \lambda (x + \Delta x) \; f(x + \Delta x) \Delta x) \; \Delta x
\end{equation}

Replacing (26), (27) and (33) into (29) gives
\begin{equation}
(x + \Delta x) f(x + \Delta x) \Delta x = x f(x) \Delta x +(p f(x) \Delta x - \lambda (x + \Delta x) f(x + \Delta x) \Delta x) \Delta x
\end{equation}

Simplifying one $\Delta x$ and transposing the first term in the right hand side to the left
\begin{equation}
(x + \Delta x) f(x + \Delta x) - x f(x) = p f(x) \Delta x - \lambda (x + \Delta x) f(x + \Delta x) \Delta x
\nonumber
\end{equation}

The first term is expanded to
\begin{equation}
x f(x + \Delta x) - x f(x) + \Delta x f(x + \Delta x) 
\nonumber
\end{equation}

Dividing both sides by $(x \Delta x)$
\begin{equation}
\frac{f(x+\Delta x)-f(x)}{\Delta x} +\frac{f(x+\Delta x)}{x} =\frac{p}{x}
f(x)-\lambda \cdot f(x+\Delta x)-\frac{\lambda \cdot \Delta x\cdot
f(x+\Delta x)}{x} 
\nonumber
\end{equation}

Finally, taking the limit when 

$\Delta x\rightarrow 0$
 we obtain the differential equation

\begin{equation}
\frac{df(x)}{dx} =(p-1)\frac{f(x)}{x} -\lambda \cdot f(x)
\end{equation}

or
\begin{equation}
\frac{f'(x)}{f(x)} =\frac{(p-1)}{x} -\lambda 
\end{equation}

whose solution is
\begin{equation}
f(x)=c\cdot x^{p-1} e^{-\lambda \cdot x} 
\end{equation}

where \textit{c} is some constant of integration that is computed 
integrating the distribution and normalizing to 1 and whose result is
\begin{equation}
c=\frac{\lambda ^{p} }{\Gamma (p)}
\end{equation}

The distribution thus obtained, \textit{f}(\textit{x}), is the powerful Gamma distribution. \\

If the value of the variable \textit{p} is particularized to an integer value then this distribution converts into the Erlang or Poisson distributions. \\

If the variable \textit{p} has the value 1, the Gamma distribution converts into the negative exponential distribution, also called Boltzmann, Gibbs, Boltzmann-Gibbs or simply exponential law.\\

In order to have an expression without the exponential but only 
with the power term we need that the parameter $\lambda$ be zero. But, in that case, it can be proved that this distribution 
does not converge.

However, in the area of econophysics it is very common the use 
of the so-called Pareto or power law distribution, although it 
is obvious that it is not a probability distribution \cite{pareto}. It should 
be profitably replaced using the Gamma distribution with the proper 
parameters.

If a quantity \textit{x} is distributed according to the Gamma distribution then its average is
\begin{equation}
\left\langle x\right\rangle =\frac{p}{\lambda }
\end{equation}

As a result, if we keep the average of \textit{x} equal to the unity then \textit{p} must be equal to $\lambda $. Nevertheless the larger the value of $\lambda $ the smaller the variance since this is given by:
\begin{equation}
\sigma ^{2} =\frac{p}{\lambda ^{2} } 
\end{equation}

The maximum of the Gamma distribution is at the position

\begin{equation}
x_{\max } =\frac{p-1}{\lambda } =\left\langle x\right\rangle
-\frac{1}{\lambda } 
\end{equation}

This means that the maximum is always towards the left of the 
average, except in the limit when $\lambda \rightarrow \infty $.

The parameters $\lambda $ and \textit{p} used in this distribution should have in general 
values greater than or equal to 1. Accordingly, if the saving propensity 
of the agents is presupposed to be some variable between zero 
and 1 we will have to compute its inverse in order to get the 
needed parameter $\lambda $. The maximum is displaced toward the left of the average precisely 
in proportion to the saving propensity.\\

``The functional form of such a distribution has been conjectured to be a Gamma distribution on 
the base of an analogy with the kinetic theory of gases, which 
is consistent with the excellent fitting provided to numerical 
data'' \cite{patriarca0},\cite{patriarca1},\cite{patriarca2}.\\

\textbf{6. ENERGY DISTRIBUTION OF AN IDEAL GAS}\\

The Maxwell equation describes the energy distribution of the 
molecules of an ideal gas. This was one of the first applications 
of statistical methods in Physics. James C. Maxwell obtained 
it for the first time around 1857. It has the form \cite{alonso} \begin{equation}
\frac{dn}{dE} =\frac{2\pi N}{(\pi \cdot kT)^{\frac{3}{2} } }
\frac{E^{\frac{1}{2} } }{Exp(\frac{E}{kT} )} 
\end{equation}

where \textit{dn}/\textit{N} is the proportion of molecules with energy 
between \textit{E} and \textit{E}+\textit{dE}. 

The Gamma distribution can be defined as the following function
\begin{equation}
P(p,\lambda ,x)=\frac{\lambda ^{p} \cdot x^{p-1} }{Exp(\lambda \cdot x)\cdot \Gamma (p)} 
\end{equation}

We recover the Maxwell equation after replacing the parameters of the Gamma distribution with the values \textit{p}=\textit{3/2} and $\lambda $
=1/\textit{kT}:

\begin{equation}
P(\frac{3}{2} ,\frac{1}{kT} ,E)
\end{equation}

With these values we obtain the average energy of the particles 
\cite{alonso}:

\begin{equation}
\langle E \rangle =\frac{p}{\lambda} = 3 kT/2
\end{equation}

This means that the Maxwell equation has simply been a Gamma distribution with some absolutely specific parameters (\textit{p}=\textit{3/2}, $\lambda $
=1/\textit{kT}).\\

In the equation for energy distribution it is possible to replace the energy variable everywhere by its equivalent kinetic energy, in terms of velocity squared. The definition suggested above provides a solid mathematical frame of reference for not doing that in an ad hoc fashion. The distribution of speeds becomes:

\begin{equation}
\frac{dn}{dv} =\sqrt{\frac{2}{\pi } } N\left( \frac{m}{kT} \right)
^{\frac{3}{2} } \frac{v^{2} }{Exp(\frac{m\cdot v^{2} }{2kT} )} 
\end{equation}

And the average of \textit{v}$^{2}$ is obtained from the formula for average 
energy, replacing \textit{E} by the kinetic energy:

\begin{equation}
\langle \frac{1}{2} \textit{m v}^{2}\rangle =p/ \lambda = 3kT/2
\end{equation}
or
\begin{equation}
\langle v^{2}\rangle = 3kT/m
\end{equation}

\textbf{7. AN APPLICATION OF THE GAMMA DISTRIBUTION}\\

No matter what specific values we assign to the parameters ``\textit{p}'' and ``$\lambda$'', they always produce a 
probability distribution (integral equal to unity) so the 
values selected for a given application must be given experimentally or by other means. \\

In the previous section, for example, we used the particular values \textit{p}=\textit{3/2} and $\lambda $
=1/\textit{kT}, with which we were able to reproduce the Maxwell equation. In the following we will try to find the appropriate values to describe the blackbody radiation distribution.\\

The first alternative is to try to obtain the exponent 3 in the 
variable $\nu $ of the Gamma distribution, in order that it be equal to the corresponding term in the Planck distribution expressed in terms of frequencies. This gives the value 4 for the parameter \textit{p}, which has the support of Wien's analysis. 

As was explained for equation (39), the value of \textit{p} fixes 
automatically the value of the variable $\lambda $ to the same value 4 since we can assume that the average frequency 
is the unity.\\

In such case, the equivalent to the Stefan-Boltzmann constant becomes:
\begin{equation}
\sigma =\frac{4^{4}\cdot k^{4}}{6\cdot h^{3}\cdot c^{2}} 
\end{equation}\\
Whose numerical value is: $\sigma = 5.927\cdot 10^{-8} \cdot [W m^{-2} K^{-4}] $, given the current value of \textit{k}, and is close to the current value: $\sigma = 5.668\cdot 10^{-8} \cdot [W m^{-2} K^{-4}]$.\\

Normalizing the Planck distribution to unit area becomes:
\begin{equation}
\frac{15p^{4} x^{p-1} }{\pi ^{4} (e^{p\cdot x} -1)} 
\end{equation}
It is illustrative to compare this Planck distribution (with \textit{p}=4) 
with the Gamma distribution \textit{P}(4,4,\textit{x}), in Figure 1.
\begin{figure}[hbtp]
\begin{center}
\includegraphics[viewport=-50 0 600 320,width=15cm,clip]
{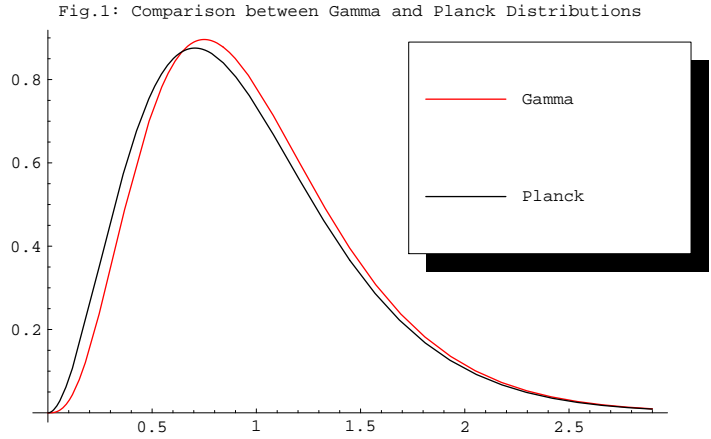} 
\caption{Gamma vs. Planck}
\end{center}
\end{figure}

Although both distributions are close it must be noted that they 
are different. The maximum for the Gamma distribution is exactly 
at \textit{x}=0.75 whereas the maximum for the Planck distribution 
is at \textit{x}=0.70536. The maximum values are respectively 0.896167 
and 0.875546, with around 2.3\% percent difference between them.\\

Now let us try to fit both the Planck and Gamma distributions 
to the experimental data from the cosmic background radiation, 
collected by the COBE satellite \cite{smoot}, \cite{cobefiras}.\\

The plot of the COBE data (blue dots) together with the Planck distribution with parameter \textit{p}=4 and 
scale factors vertical=437.987 and horizontal=7.72655 (green), 
and Gamma distribution with parameters \textit{p}= $\lambda $=4 and scale factors vertical=427.909 and horizontal=7.26667 
(red), is shown in Figure 2. The adjustment 
was made around the point of maximum radiation.
\begin{figure}[htbp]
\begin{center}
\includegraphics[viewport=-50 0 600 320,width=15cm,clip]
{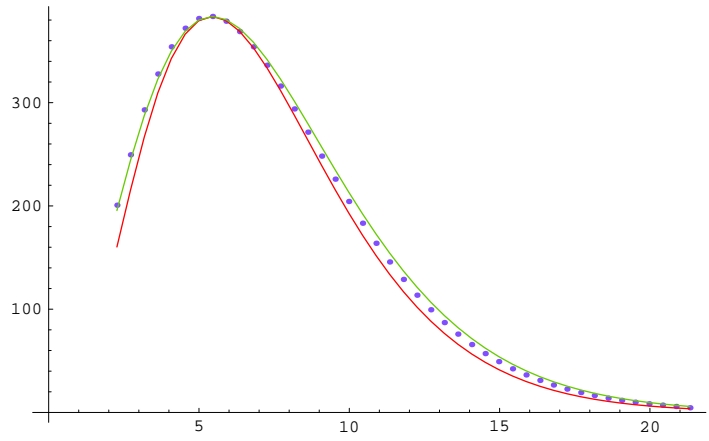} 
\caption{COBE data}
\end{center}
\end{figure}

Although there is a good fit of both curves there are some zones 
where we can note some discrepancy. \\

Just to experiment I made the fit of a Gamma distribution with 
parameters p=$\lambda $=3.5 and scale factors vertical=451.258 and horizontal=7.36276 that gives a visually almost perfect fit (Figure 3). \\
\begin{figure}[hbtp]
\begin{center}
\includegraphics[viewport=-50 0 600 320,width=15cm,clip]
{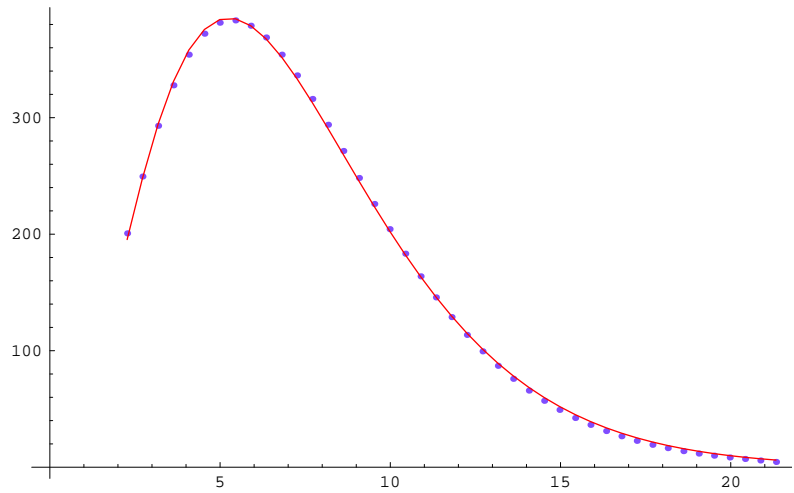} 
\caption{COBE data (p=3.5)}
\end{center}
\end{figure}

The figures suggest that neither the Planck distribution nor 
the Gamma with parameters p= $\lambda $=4 are perfect to adjust the COBE data. The Gamma distribution with p= $\lambda $=3.5 is much better; however, it is necessary 
to analyze more results in order to confirm or refute the possibility to change the parameter \textit{p} to the new value 3.5, or to some other value.\\

In the same sense that the Gamma distribution integrates to the 
unity for any combination of ``p'' and ``$\lambda $'', both forms of the Planck distribution integrate to the same 
(Stefan-Boltzmann) expression. But this does not mean that both 
equations are valid, as it is not true that any form of the Gamma 
distribution is adequate for expressing the energy distribution 
of gases, photons, money, etc. We have to find the correct values 
for ``p'' and ``$\lambda $''.\\

In general, ``p'' depends on the degrees of freedom and ``$\lambda $'' depends on the kinetic energy of the system. \\

\textbf{8. CONCLUSIONS}\\

It is experimentally found that a black body produces a radiation distribution that has a given maximum density. Associated with this maximum there must be some kind of photons that carry the maximum amount of energy per unit time (or power). Those photons should have some precise photon frequency; however let us assume that such frequency is not specified at the moment.\\

In particular, the current blackbody radiation theory seems to lack an explanation as to what the (frequency or wavelength) characteristic of this kind of photons is. \\

This paper evidences that two different maxima, with two different kinds of photons, are computed, corresponding to the two Planck's equations. One of the equations describes the blackbody radiation energy distribution per unit wavelength, expressed as a function of wavelength, and the other equation describes the blackbody radiation energy distribution per unit frequency, as a function of frequency.\\

The variables \textit{maxnu} and \textit{maxlambda}, defined in section 2, can be used to compute the values for the two corresponding kinds of photons.\\

The two different kinds of photons proposed by current physics seem to be just mathematical values necessary to compute the said maxima for the two corresponding Planck's radiation distribution densities, but it is not proved that either one of these values necessarily corresponds to the photons which carry the maximum radiation power.\\

In this paper it was proved that if we assimilate the wealth of the persons in a population to continuous photon energy, then the probability distribution, for a population in thermodynamic equilibrium, of the number of persons with increasing wealth, seems to follow a Gamma distribution. Now, if we multiply the number of persons with a given wealth by its corresponding wealth we obtain the amount of money held by that population group.\\

Similarly, the number of photons produced by blackbody radiation seems to follow another Gamma distribution. If we multiply the number of photons carrying a given energy by its corresponding energy (or frequency by Planck's constant) we obtain the amount of energy held by that kind of photons (assume per unit of time).\\

Consequently, if the temperature variable of equation (18) is replaced by the corresponding photon frequencies this should provide the actual power distribution.\\

It should be realized that the quotient of the speed of light divided by the photon wavelength is the \textit{same} variable as photon frequency; this means that there should be only one probability distribution. In other words, there is no way for frequency and wavelength to be interpreted as independent variables with different and independent probability distributions. \\

In current Physics now and always remains the same centuries-old spirit that nothing can be advanced in Physics and that the old theories are correct. Fortunately 
there always are some stubborn people that do not believe in 
dogmas and tries to understand and explain the behavior of Nature. 
For example, it is known that Planck himself ``was attracted 
to Physics despite having been advised that nothing essentially 
remained to be discussed in this branch of learning'' \cite{kamble} or that 
``in this field, almost everything is already discovered, and 
all that remains is to fill a few holes.'' \cite{wiki}\\

Although a good fit to the Cosmic Background Radiation data was shown in this paper, it should be taken with skepticism. For the previous version of the present paper I was convinced that the COBE experiment was a good testbed for my equations, but later I investigated a little more and found that such experiment assumed the prior validity of the Planck distribution and only tried to detect discrepancies with respect to such distribution. As explains Dr. Dale Fixsen, former director of the COBE project, in a private communication: ``The FIRAS measurement was a differential measurement. The measurement was not the spectrum of the CMBR but the difference between the CMBR spectrum and the blackbody spectrum provided by the on-board external calibrator. The display of the spectrum implicitly assumes the Planck spectrum is a good description of the blackbody spectrum.'' \cite{fixsen}. Also, Dr. John Mather explained to me that ``the COBE results are based on assuming that the Planck formula is correct, so you can not logically use the COBE results to test your theory.'' \cite{mather}.

Even though such project cannot be blamed for having used what at the moment could have appeared as the correct theory, I expect that new experiments be carried out to determine whether the here proposed distribution is a better physical model.\\

The Gamma distribution is an all pervasive function that, although without any theoretical support, has 
been applied successfully in numerical applications such as the 
distribution of money among several agents (econophysics) or 
could profitably be applied to describe the energy and speed distributions in ideal and real gases as a function of temperature (or, better, as a function of frequency), distribution of population in cities, etc. While the origin of the Gamma distribution 
involves continuous variables its range of applicability should 
not be strictly limited to them.

In relation to its practical use, Wien's equation is a very good one. As a Physics teacher says: ``Wien's law does not look too bad either, we can calculate an approximation for these constants A and $\beta$, and the new formula fits everywhere'' \cite{born}.\\

We have to distinguish among the new form for entropy, proposed in the present paper as the incomplete Gamma function, and the probabilistic Gamma distribution, which is obtained by the derivative of the incomplete Gamma function.
 
Note that the Euler's Gamma function satisfies:

\begin{equation}
\Gamma (p)=\int\nolimits_{0}^{\infty }x^{p-1} e^{-x} dx 
\end{equation}

This expression is the equivalent, for the Gamma distribution, of the \textbf{partition function} defined in classical thermodynamics for the Boltzmann distribution.\\


\begin{thebibliography}{99}
\bibitem{blumler}
Blumler, P. ``Thermal and Statistical Physics'', page 88. School of Physical Sciences, University of Kent at Canterbury. (Already disappeared from the Internet. I wonder if it is the problem of the establishment rejecting the messenger instead of the message. In any case, I am including new alternative references in this bibliography).
\end{thebibliography}
\end{document}